\documentclass[preprint,showpacs,aps]{revtex4}
\usepackage{graphicx}
\usepackage{graphicx}
\usepackage{amsmath,amssymb}
\begin{document}

\title{Spatially confined Bloch oscillations in semiconductor superlattices}

\author{ L. L. Bonilla, M. \'Alvaro, M. Carretero}
\affiliation{
Gregorio Mill\'an Institute for Fluid Dynamics, Nanoscience and Industrial Mathematics, Universidad Carlos III de Madrid,
Avenida de la Universidad 30, 28911 Legan\'es, Spain, EU}
\begin{abstract}
In a semiconductor superlattice with long scattering times, damping of Bloch oscillations due to scattering is so small that convective nonlinearities may compensate it and Bloch oscillations persist even in the hydrodynamic regime. In this case, numerical solutions show that there are stable Bloch oscillations confined to a region near the collector with inhomogeneous field, charge, current density and energy density profiles. These Bloch oscillations disappear when damping due to inelastic collisions becomes sufficiently strong. 
\end{abstract}


\maketitle

\section{Introduction}
Bloch oscillations (BOs) are coherent oscillations of the position of electrons inside energy bands of a crystal under an applied constant electric field $-F$. Their frequency is $\omega_B =eFl/\hbar$ ($l$ lattice constant), and therefore it can be tuned by an applied voltage. BOs were predicted by Zener in 1934 as an immediate consequence of the Bloch theorem \cite{zener}, but they were not observed for almost sixty years because scattering of the electrons with phonons, impurities, etc.\ damp them very rapidly into oblivion. To observe BOs, their period has to be shorter than the scattering time $\tau$, and therefore the applied field has to surpass the value $\hbar/(el\tau)$, which is too large for most natural materials, in which $l$ is of \AA ngstr\"om size. In 1970, Esaki and Tsu suggested to create an artificial crystal, which they called superlattice (SL), by growing many identical periods comprising a number of layers of two different semiconductors with similar lattice constants \cite{esaki}. The period of the resulting one-dimensional crystal may be much larger, say about 10 nm, and this gives reasonable electric fields of about 10 kV/cm, which are within the range of experimental observation. Damped Bloch oscillations were first observed in 1992 in semiconductor SLs whose initial state was prepared  optically \cite{fel92}. Besides their interest for theoretical physics, BOs have attracted the attention of many physicists and engineers because of their potential for designing infrared detectors, emitters or lasers which can be tuned in the THz frequency range simply by varying the applied electric field \cite{leo}. However no electrically driven devices based on BOs have been realized. Another application is based on the fact that BOs give rise to a resonance peak in the absorption coefficient under dc+ac bias and a positive gain at THz frequencies \cite{ktitorov}. The latter has been observed in quantum cascade laser structures \cite{ter07}. These applications are severely limited by scattering which rapidly damps BOs and, for a dc voltage biased SL, favors the formation of electric field domains (EFDs) whose dynamics yields self-sustained oscillations of lower frequency (GHz) \cite{hof96,BGr05} (a phenomenon similar to the Gunn effect in bulk GaAs \cite{kroemer}). EFD formation may also preclude THz gain in simple dc+ac driven SL which is typically calculated assuming spatially uniform solutions of drift-diffusion or Boltzmann type equations \cite{leo,kro00a,kro00b,ale06,hya09,IRo76}. This assumption has not been tested by solving space-dependent equations with appropriate boundary conditions or by experiments in semiconductor superlattices. An interesting idea for efficient terahertz harmonics generation is to excite relaxation oscillations in the superlattice by incident radiation from a waveguide \cite{ign11}.

To understand the role of EFD formation in the observation of BOs or THz Bloch gain, our starting point should be a model in which BOs and EFDs are both possible solutions of the governing equations. One simple possibility is to use a self-consistent version of the Ktitorov, Simin and Sindalovskii (KSS) Boltzmann equation \cite{ktitorov} with Bhatnagar-Gross-Krook (BGK) collision terms \cite{ISh87,BEP03}. The characteristic equations of this kinetic equation exhibit BOs as solutions, whereas there exist a hydrodynamic regime for large applied electric fields that yields Gunn-type oscillations of the current as solutions of a drift-diffusion equation \cite{BEP03}. However the KSS model does not have a hydrodynamic regime of the kinetic equation in which both BOs and EFDs are possible solutions. Why is the coexistence of BOs and EFDs not possible in the KSS model? Firstly, the BOs and the hydrodynamic regime correspond to widely separated time scales. In the slower hydrodynamic regime, BOs have already disappeared due to scattering and we cannot study simultaneously BOs and the EFDs appearing in the hydrodynamic regime. This problem could be reduced if we consider materials with long-lived BOs corresponding to almost elastic collisions, as will be discussed in more detail later. The second problem is that, unlike the original BGK local equilibrium \cite{BGK}, the local equilibria used in \cite{ISh87} and in \cite{BEP03} depend only on the 2D electron density $n$, which remains approximately constant during a BO. Since hydrodynamic regimes are perturbations of local equilibria, the KSS model as modified in \cite{BEP03} can only provide slowly varying drift-diffusion equations \cite{footnote1} for the electron density and the electric field $-F$ that cannot contain BOs among their solutions. This second problem can be solved if the local equilibrium in the Boltzmann-BGK kinetic theory depends on electron density, electron current density and mean energy and the collision term preserves charge but dissipates momentum and energy \cite{BC08}. {\em The most important property of the proposed model is that it allows the local equilibrium distribution to oscillate at the Bloch frequency,} which is the crucial feature (missing in the KSS kinetic equation) if we want to derive a hydrodynamic regime that allows BOs. 
\section{Model}
The model equations are
\begin{eqnarray}
&&\partial_{t} f + v(k)\, \partial_{x} f +  e F\hbar^{-1} \partial_{k} f =
Q[f]\equiv - \nu (f - f^{B}),\quad  \label{1}\\
&&\varepsilon\, \partial_{x} F = e l^{-1} (n-N_{D}),  \label{2}
\end{eqnarray}
\begin{eqnarray}
&& f^{B}(k;n,J_{n},E) = n\,\frac{\pi\, e^{ \tilde{u}kl +\tilde{\beta}\cos kl}}{
\int_{0}^\pi e^{\tilde{\beta}\cos K}\cosh(\tilde{u}K)\, dK}, \label{3}
\end{eqnarray}
\begin{eqnarray}
 n = { l\over 2\pi} \int_{-\pi/l}^{\pi/l} f(x,k,t) dk = { l\over
2\pi} \int_{-\pi/l}^{\pi/l} f^{B} dk.    \label{4}
\end{eqnarray}
Here $n$, $N_{D}$, $\varepsilon$, $-e<0$, $m^*$, $\nu$, and $-F$ are the 2D electron density, the 2D doping density, the permittivity, the electron charge, the effective mass of the electron, the constant collision frequency and the electric field, respectively. $v(k)=\Delta l\sin(kl)/(2\hbar)$ is the group velocity corresponding to the miniband tight binding dispersion relation $\mathcal{E}(k)=\Delta(1-\cos kl)/2$. For the sake of simplicity, we have assumed a Boltzmann local equilibrium (\ref{3}), but it is easy to replace it by the Fermi-Dirac local distribution in the degenerate case. The distribution functions $f$ and $f^B$ have the same units as $n$ and are $2\pi/l$-periodic in $k$ (the function $\tilde{u}kl$ in (\ref{3}) is extended periodically outside $-\pi<kl\leq \pi$). A quantum version of (\ref{1}) can be obtained as indicated in Ref.\ \cite{BGr05}.

The dimensionless multipliers $\tilde{\beta}(x,t)$ and $\tilde{u}(x,t)$ depend on $J_{n}$=$e \int_{-\pi/l}^{\pi/l} v(k)\, f\, dk/(2\pi)$ (electron current density) and on $E$=$l \int_{-\pi/l}^{\pi/l}[\Delta/2- \mathcal{E}(k)] f dk/(2\pi n)$ (mean energy). They are found by solving
\begin{eqnarray}
&& {e\over 2\pi} \int_{-\pi/l}^{\pi/l} v(k)\, f^{B}\, dk= (1-\alpha_{j})J_{n},
\nonumber\\
&& {l\over 2\pi n} \int_{-\pi/l}^{\pi/l}  \left({\Delta\over 2} -{\cal E}\right)
f^{B} dk = \alpha_{e} E_{0} + (1-\alpha_{e}) E.\quad \label{5}
\end{eqnarray}
Similarly to BGK collision models in rapid granular flows \cite{BMD}, the restitution coefficients $\alpha_{j}$ and $\alpha_{e}$ take values on the interval $[0,1]$ and measure the dissipation due to collisions in current density and energy, respectively. In fact, the collision operator satisfies $\int_{-\pi/l}^{\pi/l} Q[f]\, dk = 0$ (charge continuity),  $e\int_{-\pi/l}^{\pi/l} v(k)\, Q[f]\, dk/(2\pi) = -\nu\alpha_{j} J_{n}$,  and $l\int_{-\pi/l}^{\pi/l} [\Delta/2-{\cal E}(k)]\, Q[f]\, dk/(2\pi n) = -\nu\alpha_{e} (E-E_{0})$. Obviously for $\alpha_{e,j}=0$ the collisions conserve energy and momentum (elastic limit). To simplify matters, we shall assume that $\alpha_{j}$ and $\alpha_{e}$ are constant. $E_{0}$ is the mean energy at the lattice temperature of the global equilibrium which will be reached in the absence of bias and contact with external reservoirs. Equivalent results are obtained if we define the mean energy as the average of ${\cal E}(k)$, which is equal to $\Delta/2 - E$, but Eq.\ (\ref{5}) leads to a simpler relation between energy density and lattice temperature. At the lattice temperature, $T_{0}=\Delta/(2k_{B}\tilde{\beta}_{0})$, $\tilde{u}=0$, $E=E_{0}$, and (\ref{5}) yields $2E_{0}/\Delta= I_{1}(\tilde{\beta}_{0})/I_{0}(\tilde{\beta}_{0})$, where $I_{s}(x)$, $s=0,1$, are modified Bessel functions.

\section{Hydrodynamic equations} 
In the hyperbolic limit in which the collision and Bloch frequencies are comparable and dominate all other terms in (\ref{1}), it is possible to derive closed equations for nondimensional $n$, $F$ and $A$ (the complex envelope of the BO solution) \cite{derive},
\begin{eqnarray}
f_1=A(x,t) e^{-i\theta}+f_{1,S}(x,t),\quad 
\theta=\frac{1}{\delta} \int_0^t F(x,s)\, ds \label{sl0}
\end{eqnarray}
($\theta$ is the rapidly varying phase of the BO, $f_{1,S}=O(\delta)$ is written below), provided the collisions are almost elastic. The small dimensionless parameter $\delta=e^2N_D l\Delta/(2\varepsilon\hbar^2\nu^2)$ is the ratio between the scattering time and the dielectric relaxation time and the restitution coefficients are assumed to scale with it, $\alpha_{e,j}=\delta\gamma_{e,j}$. A similar double limit of vanishing Knudsen number (equivalent to $\delta\to 0$) and almost elastic collisions has been used to derive hydrodynamic equations for granular gases \cite{SG98}\cite{footnote2}. In (\ref{sl0}), $f_j(x,\theta,t;\delta)$ are the Fourier coefficients of $f(x,k,t;\delta)=\sum_{j=-\infty}^\infty f_j e^{ijk}$ and $f_1=nE-iJ_n$. The nondimensional equations are
\begin{eqnarray}
\frac{\partial F}{\partial t} &+&\frac{\delta}{F^2+\delta^2\gamma_{j}
\gamma_e} \left[\gamma_{e}E_{0}nF+\frac{F}{2}\, {\partial\over
\partial x}\mbox{Im}\,\frac{f^{B(0)}_{2,0}}{1+2iF}\right.\nonumber\\
&-&  \frac{\delta\gamma_{e}}{2}\, {\partial\over
\partial x}\left(n-\mbox{Re}\,\frac{f^{B(0)}_{2,0}}{1+2iF}\right) -
F \mbox{Re}\, h_{S}\nonumber\\
&+& \left.\delta
\gamma_{e}\mbox{Im}\, h_{S}\right] =J(t),   \label{sl1}\\
\frac{\partial F}{\partial x} &=&n-1,  \label{sl2}\\
\frac{\partial A}{\partial t} &=&-\frac{\gamma_{e}+\gamma_{j}}{2}
A + \frac{1}{2i}\frac{\partial}{\partial x}\left(\frac{f^{B(0)}_{2,-1}}{1+iF}\right),  \label{sl3}
\end{eqnarray}
\begin{eqnarray}
&& h_S=\frac{f_{1,Su}}{n}\frac{\partial \mbox{Im}f_{1,Su}}{\partial x}+(J+\mbox{Im}f_{1,Su})\,\frac{\partial f_{1,Su}}{\partial F}, \label{sl6}\\
&& f_{1,Su}= \frac{\delta\gamma_{e}nE_{0}(\delta
\gamma_{j}-iF)}{\delta^2\gamma_{e}\gamma_{j}+ F^2}, \nonumber
\end{eqnarray}
\begin{eqnarray}
f_{1,S} &=& nE_{S}-iJ_{n,S}\nonumber\\
 &=&\frac{\delta}{F^2+\delta^2\gamma_{j}\gamma_e}
\left[\gamma_{e}nE_{0} (\delta\gamma_{j}-iF)\right.\nonumber\\
&-&(\delta\gamma_{j}-iF)\mbox{Re}h_{S}-(F+i\delta\gamma_{e})\mbox{Im}h_{S}\nonumber\\
&+& \frac{F+i\delta\gamma_{e}}{2}\,
\frac{\partial}{\partial x}\left(n-\mbox{Re}\, \frac{f_{2,0}^{B(0)}}{1+i2F}\right)  \nonumber\\
&+& \left.\frac{\delta\gamma_{j}-iF}{2}\, \mbox{Im}\frac{\partial}{\partial x} \left(\frac{f_{2,0}^{B(0)}}{1+i2F}\right)  \right]. \label{sl8}
\end{eqnarray}
We have defined the nondimensional variables $\tilde{f}= f/N_D$, $\tilde{n}=n/N_D$, $\tilde{E}=2E/\Delta$, $\tilde{J}_n=J/[J_n]$, $\tilde{x}=x/[x]$, \ldots (where $[y]$ are the units in Table \ref{t1}) and omitted tildes over variables. The dimensionless multipliers $\tilde{\beta}$ and $\tilde{u}$ in $f^B$ are functions of the rapidly varying BO phase $\theta$ due to (\ref{sl0}) and therefore, we can expand $f^B$ in (\ref{3}) in powers of $\delta$, $f^B\sim f^{B(0)} +\delta f^{B(1)}$. The $f^{B(m)}$ ($m=1,2)$ are now $2\pi$-periodic functions of $\theta$ and $k$. Then we have the Fourier coefficients
\begin{eqnarray}
f^{B(0)}_{j,m} &=& \int_{-\pi}^\pi \int_{-\pi}^\pi f^{B(0)}(k;n,f_1)\, e^{-ijk-im\theta}\frac{dk\, d\theta}{(2\pi)^2} , \label{sl7}
\end{eqnarray}
in which we set $f_1=A\, e^{-i\theta}$ ignoring $O(\delta)$ terms in (\ref{sl0}). 

To derive (\ref{sl1})-(\ref{sl3}), we start from the equations for the moments $f_j$ which can be obtained from (\ref{1}) by integration over $k$ \cite{BC08}:
\begin{eqnarray} 
&& \frac{\partial f_0}{\partial t} - \mbox{Im}\frac{\partial f_{1}}{\partial x}=0,  \label{sl7.1}\\
&&\left(\delta\frac{\partial }{\partial t}+iF\right)f_1= \delta\left[\gamma_ef_0E_0-\frac{\gamma_e+\gamma_j}{2}f_1-\frac{\gamma_e-\gamma_j}{2}f_1^*-\frac{1}{2i}\frac{\partial}{\partial x}(f_0- f_2)\right],  \label{s7.2}
\end{eqnarray} 
where $f_0=n$, $f_1= nE-iJ_n$, and there are similar equations for higher moments. From (\ref{sl7.1}) and the Poisson equation $\partial F/\partial x= n-1$, we find Amp\`ere's law for $F$: $\partial F/\partial t= J(t)-J_n$, where $J(t)$ is the total current density. We shall assume that the second moment $f_2$ is a known function of $f_0$ and $f_1$, $f_2=g(f_0,f_1)$. Then we find equations for $n$, $F$ and $A$ in (\ref{sl0}) by a method of nonlinear multiple scales \cite{KCo96} with time scales $\theta$ and $t$. To obtain the function $g$, we carry out a Chapman-Enskog expansion \cite{BEP03,derive} for (\ref{1}) (in dimensionless units) with a time derivative given by $F\,\partial f/\partial\theta+\delta\,\partial f/\partial t$, according to (\ref{sl0}). The distribution function is supposed to be periodic in $k$ and in $\theta$. This procedure gives approximate formulas for $g=f_2$ from which (\ref{sl1})-(\ref{sl3}) are obtained \cite{derive}.

The hydrodynamic equations (\ref{sl1})-(\ref{sl3}) have the spatially uniform solutions, $n=1$, $J=\delta\gamma_eE_0nF/(\delta^2\gamma_e\gamma_j+F^2)$ (in dimensional units this gives the well-known temperature dependent drift velocity \cite{ISh87}\cite{footnote3}), and $A=A_0 e^{-(\gamma_e+\gamma_j)t/2}$. Inserting the latter formula in (\ref{sl0}), we see that this corresponds to a damped BO whose amplitude relaxes to 0. Even when we manage to prepare the initial state with a coherent BO of complex amplitude $A_0$, ignoring space dependence will lead to disappearance of the BOs after a relaxation time $2/(\gamma_e+\gamma_j)$. Stabilization of the BOs may be caused only by the spatially dependent second term on the right hand side of (\ref{sl3}).

\begin{table}[ht]
\caption{Hyperbolic scaling and nondimensionalization with $\nu=10^{14}$ Hz.}
\label{t1}
\begin{center}\begin{tabular}{ccccccccc}
 \hline
$f$, $n$ & $F$ &${\cal E}$, $E$ &$v(k)$&$J_{n}$& $x$ & $k$ & $t$ & $\delta$\\
$N_{D}$ & $\frac{\hbar\nu}{el}$ & $\frac{\Delta}{2}$ & $\frac{l\Delta}{2
\hbar}$ & $\frac{eN_{D}\Delta}{2\hbar}$& $\frac{\varepsilon\hbar\nu}{
e^2N_{D}}$ & $\frac{1}{l}$ & $\frac{2\varepsilon\hbar^2\nu}{e^2N_{D}l
\Delta}$ & $\frac{e^2N_D l\Delta}{2\varepsilon\hbar^2\nu^2}$\\
$\frac{10^{10}}{\mbox{cm}^{2}}$ & $\frac{\mbox{kV}}{\mbox{cm}}$ & meV & $\frac{10^4\mbox{m}}{\mbox{s}}$ & $\frac{10^4\mbox{A}}{\mbox{cm}^2}$ & nm & $\frac{1}{\mbox{nm}}$ & ps & -- \\
$4.048$& 130 & 8 & 6.15 & 7.88 & 116 & 0.2 & 1.88& 0.0053\\
 \hline
\end{tabular}
\end{center}
\end{table}

\section{Results} We now solve numerically the hydrodynamic equations with the boundary conditions \cite{BGr05}
\begin{eqnarray}
&& \left.\frac{\partial F}{\partial t} +\sigma_0 F\right|_{x=0}=J, \quad
\left.\frac{\partial F}{\partial t} +\sigma_1 n F\right|_{x=L}=J, \label{sl17}\\
&& \frac{1}{L}\int_0^L F(x,t)\, dx = \phi, \label{sl18}\\
&& \frac{\partial A}{\partial x}= 0, \quad\mbox{at $x=0$.}\label{sl19}
\end{eqnarray}
We obtain similar numerical results with $A=0$ at $x=0$. Here $L=Nl/[x]$ and $\phi= eV/(\hbar\nu N)$ are the dimensionless SL length and average field (proportional to the applied voltage $V$), respectively. We have used contact conductivities $\sigma_{0,1}=12.1$ $(\Omega\,\mbox{m})^{-1}$ which yield dimensionless conductivities $\sigma_{0,1}=0.2$ (conductivity units are $[\sigma]= e^2N_D\Delta l/(2\hbar^2\nu)$). Initially, $F(x,0)=\phi$ and $A(x,0)=A_0$ (constant). The latter condition means that we have prepared the SL in an initial state having a coherent BO with complex amplitude $A_0$. Whether this can be achieved by optical means as in \cite{fel92} remains to be seen.

\begin{figure}
\begin{center}
\includegraphics[width=6.5cm,angle=0]{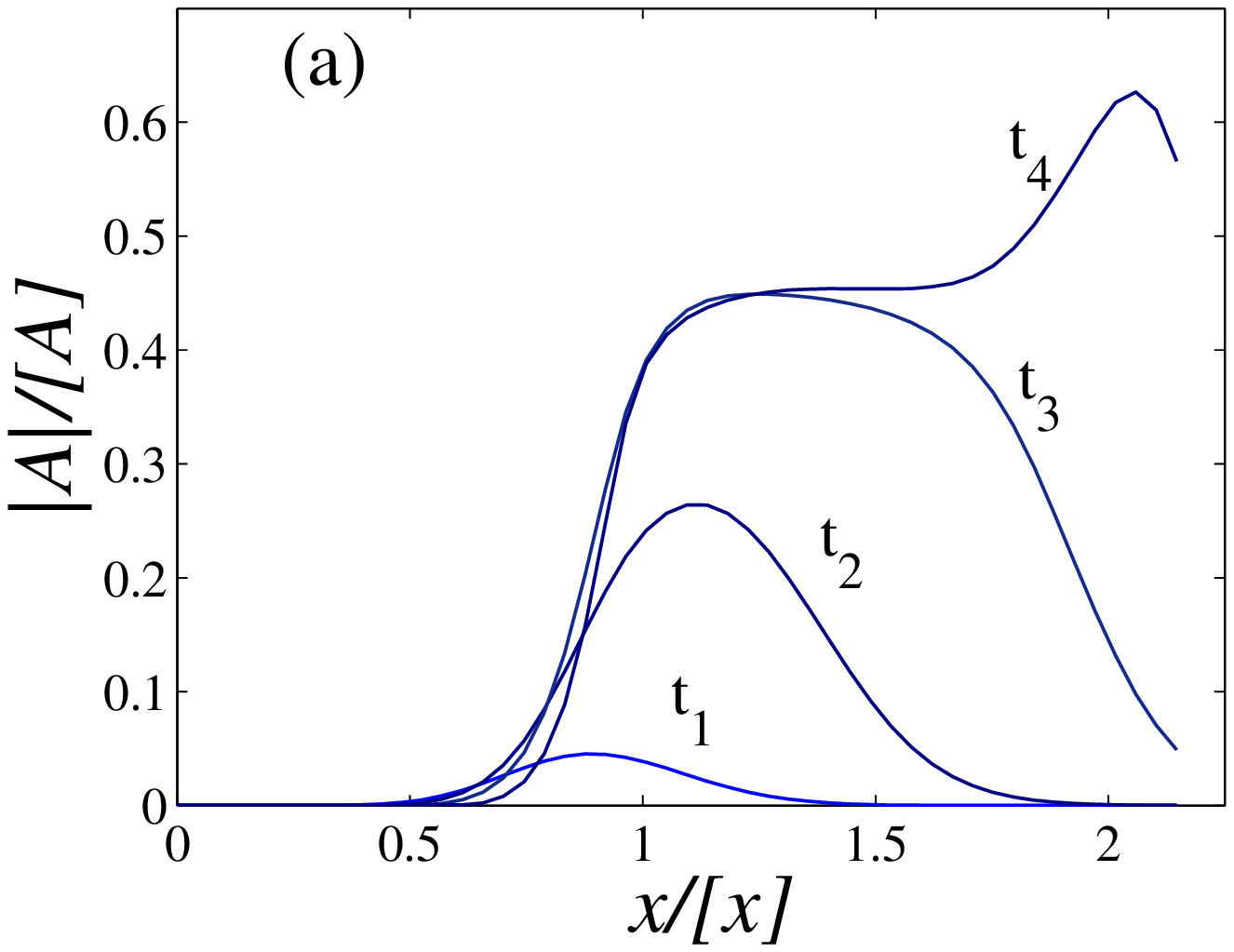}\\
\includegraphics[width=6.5cm,angle=0]{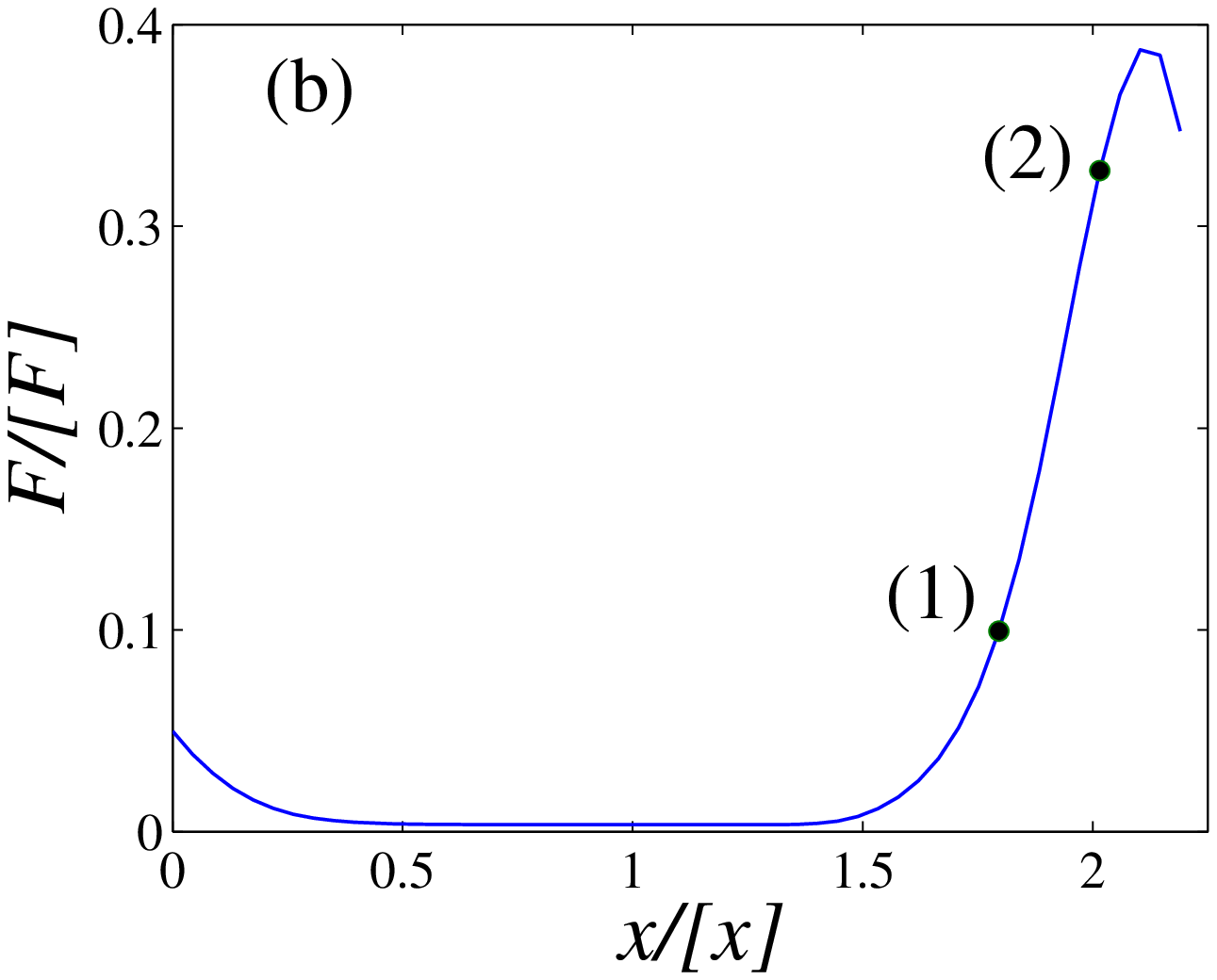}
\end{center}
\begin{center}
\includegraphics[width=6.5cm,angle=0]{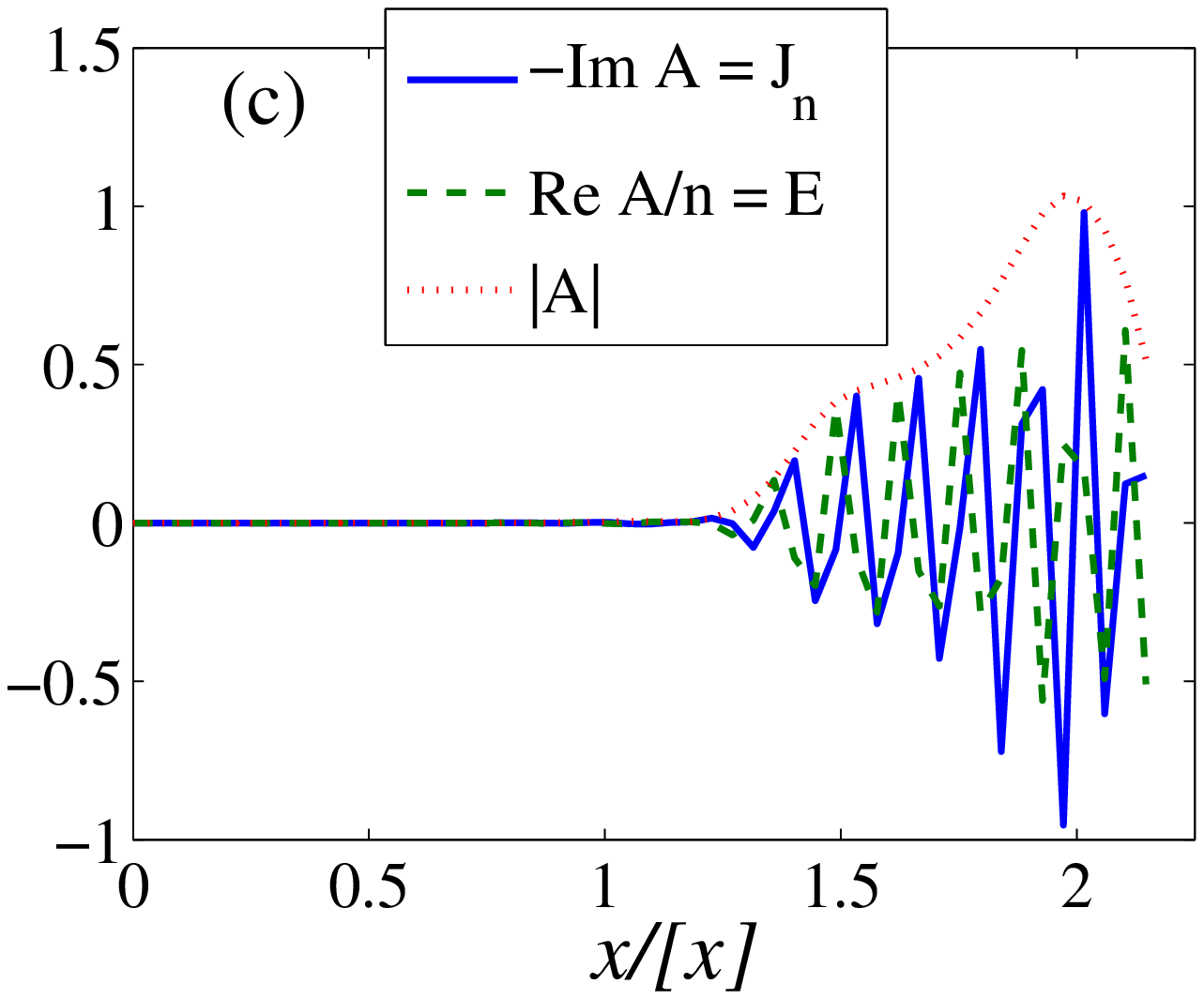}\\
\includegraphics[width=6.5cm,angle=0]{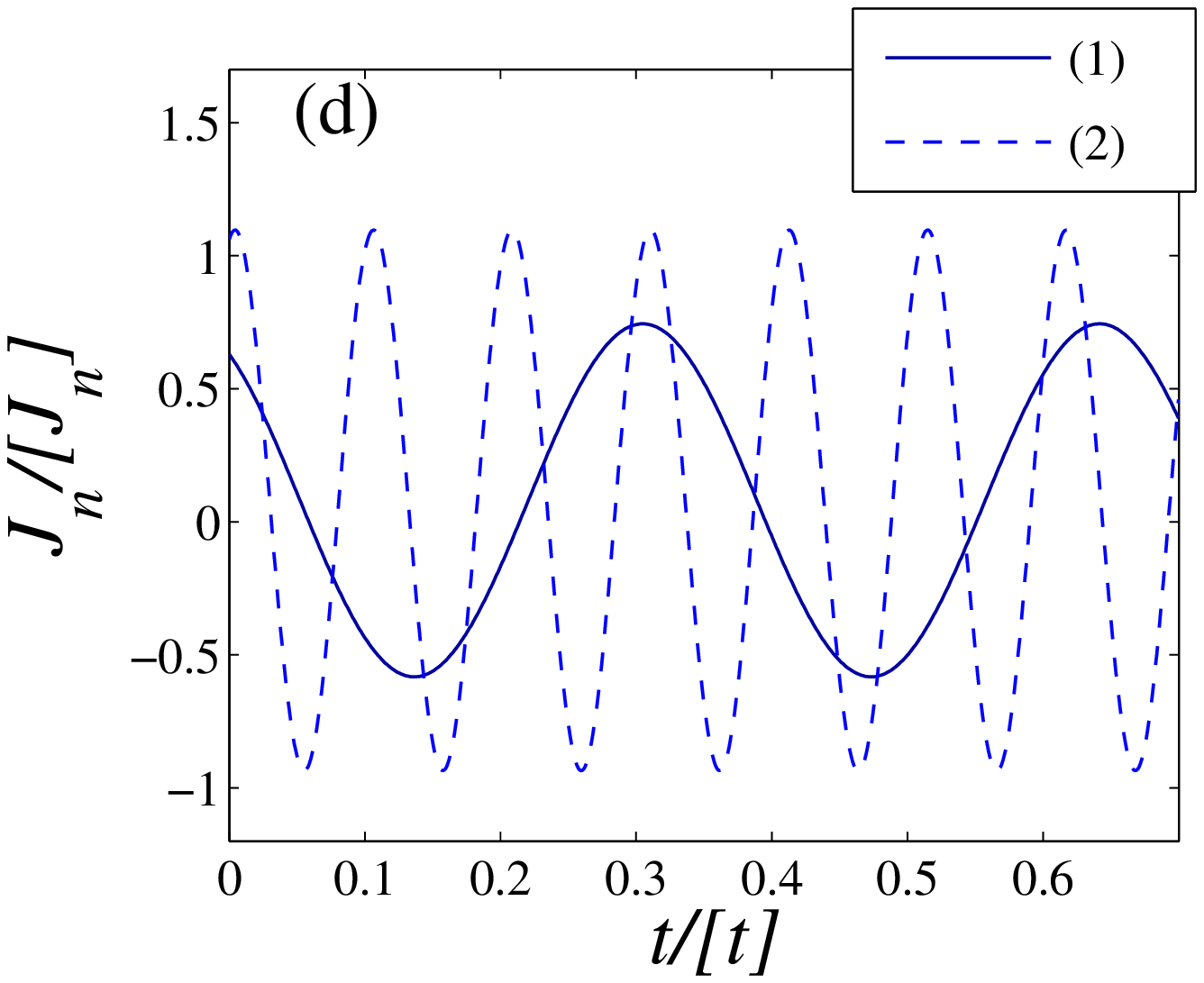}
\vspace{0.2cm} \caption{(a) Modulus of the BO complex amplitude vs space at times $t_1=7$, $t_2=9$, $t_3=11$, $t_4=13$.  (b) Stationary field profile. (c) Profiles of the nondimensional mean energy density $E$ and nondimensional electron current density $J_n$ for $\theta=0$. (d) Current density at the two different points marked by (1) and (2) in (b) during BOs. Clearly, the frequency at point (2) is larger than at (1). To transform the magnitudes in this figure to dimensional units, use Table \ref{t1}. $[A]=\Delta N_D/2$. } \label{fig1}
\end{center}
\end{figure}

We solve (\ref{sl1})-(\ref{sl3}) with the parameter values indicated in Table \ref{t1} (which are similar to those in Ref.\ \cite{sch98}) and different values of the restitution coefficients. We start with $\alpha_{e}=\alpha_{j}=0.01$ so that $\nu\alpha_{e}= \nu\alpha_j= 10^{12}$ Hz.\cite{footnote4} The 3D doping density $N_{3D}= 8\times 10^{16}$ cm$^{-3}$ gives $N_{D}=N_{3D}l=4.048\times 10^{10}$ cm$^{-2}$ as in Table \ref{t1}, and $\varepsilon=12.85\,\varepsilon_{0}$. We find $\delta\approx 0.0053$ and $\gamma_{e,j}=\alpha_{e,j}/\delta=1.8781$. We consider a 50-period ($N=50$) dc voltage biased SL with lattice temperature 300 K. For $V=0.2$ V (therefore $\phi=0.06$ \cite{footnote5}), we observe that $|A(x,t)|$ first diminishes uniformly from $A_0=0.153$ to almost zero after a relaxation time $2/(\gamma_e+\gamma_j)\approx 0.53$ (about 1 ps). Later a small pulse is formed at about $x=L/4$ which subsequently extends to the remaining part of the sample and it grows nearer to its end. The BOs are confined to the second half of the sample that is closer to $x=L$ and are zero in the first half of the sample closer to $x=0$. Thus the profile of $|A|$ has a compact support with a maximum near $x=L$. $|A(x,t)|$ is close to a  periodic oscillation in time: small pulses are formed at the left of its support, climb up towards the maximum of the pulse which then diminishes and the same behavior repeats itself. Figure \ref{fig1}(a) shows four snapshots of $|A(x,t)|$ illustrating this behavior which can also be observed in two movies \cite{footnote6}. The field profile depicted in Fig. \ref{fig1}(b) is almost stationary. The mean energy and electron current densities during one BO can be reconstructed by means of (\ref{sl0}). We show them for $\theta=0$ in Fig.\ \ref{fig1}(c). At the two different SL locations marked in Fig.\ \ref{fig1}(b), the graphs of $J_n$ versus time are shown in Fig. \ref{fig1}(d). This figure and two additional movies showing the evolution of the $J_n$ and $1-E$ profiles in the supplementary material illustrate that the Bloch frequencies depend strongly on space and are higher near the collector where the field is larger.

For the scattering times reported in Ref. \cite{sch98}, the restitution coefficients are $\alpha_e=0.09$ and $\alpha_j=0.29$, but the BO amplitude becomes zero everywhere after a short relaxation time. BOs also disappear for $\alpha_e= 0.01$ and $\alpha_j=0.29/9\approx 0.032$, and there is a smaller critical value of $\alpha_j$ (for fixed $\alpha_e$) below which BOs can be sustained. They also persist for $\alpha_j= 0.01$ and $\alpha_e= 0.09/29$ which keep the same ratio $\alpha_j/\alpha_e = 29/9\approx 3.22$ as in Ref. \cite{sch98}. There is a critical curve in the plane of restitution coefficients such that, for $(\gamma_e+\gamma_j)/2>\gamma_{\rm crit}$ ($\gamma_{\rm crit}\approx 2.5$ for $\delta= 0.0053$), BOs disappear after a relaxation time but they persist for smaller values of $(\gamma_e+\gamma_j)$.

In summary, we have analyzed the Boltzmann-BGK-Poisson equations with local equilibrium depending on the electron density, current density and energy density in the hyperbolic limit in which the BO period is much shorter than the dielectric relaxation time and collisions are almost elastic. In the long-time scale, there is a hydrodynamic regime described by coupled equations for the electric field, the electron density and the BO complex amplitude. When the restitution coefficients (equivalently the inverse of the scattering times) are sufficiently small and the initial state has been prepared so that there is a nonzero Bloch oscillation, there are stable spatially inhomogeneous profiles of current and energy densities displaying BOs confined to a fraction of the SL extent. It remains to investigate whether confined Bloch oscillations may also be important for terahertz harmonic generation with underlying inhomogeneous charge and electric field profiles.

\acknowledgments
This work has been supported by the MICINN grant FIS2008-04921-C02-01.


\begin{thebibliography}{9}
\bibitem{footnote1}
To illustrate this with an example, let us consider the simple one-dimensional problem $\partial f/\partial t+v\partial f/\partial x +F\partial f/\partial v= -(f-f^B)/\delta$ with $f^B=n e^{-v^2/2}/\sqrt{2\pi}$, $n=\int_{-\infty}^\infty f(x,v,t)\, dv$, as $\delta\to 0$. Integration over $v$ gives the continuity equation $\partial n/\partial t+ \partial J_n/\partial x=0$, with $J_n=\int_{-\infty}^\infty v f(x,v,t)\, dv$. An expansion $f\sim f^{(0)}+\delta f^{(1)}$ gives $f^{(0)}=f^B$ and $f^{(1)}= -(v\partial n/\partial x+\partial n/\partial t- nFv)e^{-v^2/2}/\sqrt{2\pi}$, which, inserted in the continuity equation, produces the simple drift-diffusion equation for $n$: $\partial n/\partial t+\delta F\partial n/\partial x=\delta\,\partial^2n/\partial x^2$. To obtain not only a continuity equation for $n$ but also equations for the average momentum and energy, we need a richer local equilibrium that depends on $n$ and also on the average momentum and energy. The original BGK local equilibrium depended on space and time-dependent density, average velocity and temperature \cite{BGK}.
\bibitem{footnote2}
The case of granular gases is simpler: there, collisions dissipate energy but conserve momentum and there is no electric field (and therefore no Bloch oscillations).
\bibitem{footnote3}
In dimensional units, the drift velocity is $v_d(F)= 2v_M\mathcal{F}/(1+\mathcal{F}^2)$, where $v_M=\Delta l I_1(\tilde{\beta}_0)/(4\hbar I_0(\tilde{\beta}_0)\tilde{\tau}_e)$, $\tilde{\tau}_e=\sqrt{\alpha_j/\alpha_e}$ and $\mathcal{F}=eFl/(\hbar\nu\alpha_e\tilde{\tau}_e)=F/F_M$. $v_M$ and $F_M$ are the velocity and field at the peak of $v_d(F)$.
\bibitem{footnote5}
This is about 6 times the peak field in the drift velocity, $F_M=\hbar\nu\sqrt{\alpha_e\alpha_j}/(el)$.
\end{thebibliography}
\end{document}